# Flaw of photonic band structure and its implication for photonic bound state in the continuum

Shahin Firuzi

*State Key Laboratory of Precision Measurement Technology and Instrument, Department of Precision Instrument, Tsinghua University, 100084 Beijing, People's Republic of China*

We show a flaw in the photonic band structure which forms the basis of our understanding of crystal momentum in photonic systems. We investigate the origin of this flaw and propose a continuous-folded band structure as the rectification. Based on the proposed continuous-folded photonic band structure, a bound state in the continuum (BIC) is actually below the light line, similar to the conventional guided modes. We present the condition for the boundedness of photonic modes and verify the effectiveness of our condition via two photonic crystal slabs with resonance-trapped BIC and symmetry-protected BIC, respectively. In the second case, our condition of boundedness correctly indicates a radiative mode, which satisfies the condition of symmetry-protected BIC.

The concept of photonic band structure was first proposed by Eli Yablonovitch [1] and Sajeev John [2] in 1987, showing that a photonic band gap similar to electronic band gap can be generated in periodic photonic structures, in which all the electromagnetic fields are suppressed. The theoretical basis and methodology of deriving the photonic band structure were developed later in 1990 [3], which became the basis for the future designs. Photonic crystals and metamaterials, whose building blocks have different optical sizes [4], both utilize the photonic band structures to analyze novel photonic phenomena such as zero refractive index [5–8] and bound state in the continuum (BIC) [9–12].

According to the conventional understanding of BICs in photonic community [12], BIC modes, unlike guided modes, are above the light line of the surrounding materials but are still confined within the photonic crystal slab. In this letter, we show that the concept of photonic band structure has a fundamental flaw in its derivation methodology [3], leading to misunderstanding of the photonic band structure, especially concepts regarding the position of the modes relative to the light line. One misunderstanding is that the BIC modes are above the light line. By clarifying the fundamental flaw of photonic band structure, we show that BIC modes are actually below the light line, and are induced by the total internal reflection, similar to the guided modes. We propose two conditions for any bound mode based on which we describe the behavior of BIC modes.

To show the connection between BIC modes and guided modes, we try to understand the boundedness due to the total internal reflection in an intuitive manner. We show that guided modes due to total internal reflection can be described, similar to BIC modes, by the destructive interference of the field at the boundary of two materials. We consider a two-dimensional model of a boundary between two materials: light propagates from the lower medium (material 1) with the refractive index $n_1$ (in this example, silicon) to the upper medium (material 2) with refractive index $n_2$ (vacuum) with an incident angle $\theta_i$ (Fig. 1). In the case that $\theta_i$ is larger than the critical angle $\theta_c = \sin^{-1}(n_2/n_1)$, total internal reflection happens so the light reflects back to the silicon slab from the boundary. The wave above the boundary is an evanescent wave. Such a wave's field distribution along the boundary consists of many pairs of fields with opposite phases (red and blue dots in Fig. 1a). For each of these pairs, the two fields interfere with each other destructively, leading to the cancellation of fields in the far-field.

The wavevector of the wave inside material 1 has a magnitude of $k_1 = n_1 \omega/c$, which can be decomposed to its components along $x$ and $y$ axes, given as $k_1^x = k_1 \sin\theta_i$ and $k_1^y = k_1 \cos\theta_i$, respectively. Under the condition of total internal reflection, along the boundary between the slab and vacuum, the $y$ component of wavevector of the superposition of incident and reflected waves becomes zero, and the wave propagates along $x$ with the wavevector $k_1^x = k_1 \sin\theta_i$. According to $k_1^x$, we can define the corresponding wavelength for the wave along the boundary as $\lambda_b = 2\pi/k_1^x$, which is the same on both sides of the boundary. By defining $\lambda_b$ and the wavelength of the light in material 2 (vacuum) as $\lambda_2$, we can write $\lambda_b = \lambda_2/\sin\theta_0$. This can be used to achieve the condition of boundedness due to the total internal reflection: when $\lambda_b < \lambda_2$, there is no real $\theta_0$ satisfying $\lambda_b = \lambda_2/\sin\theta_0$, leading to the total internal reflection (Fig. 1a). However, for $\lambda_b \geq \lambda_2$, we can find a propagation angle $\theta_0$ in material 2 to satisfy $\lambda_b = \lambda_2/\sin\theta_0$, leading to the wave propagating in material 2 (Fig. 1b).

From Fig. 1 it is clear that for a wave with $\lambda_b \geq \lambda_2$, it always leaks to medium 2. We can interpret the condition $\lambda_b < \lambda_2$ in the wavevector space as $k_1^x > k_2$. I.e. for a given frequency, if we want to bound the light inside material 1, the momentum along the boundary has to be larger than that of the surrounding material — the mode is below the light line in the band structure.

We need to consider the fact that at a flat boundary between two homogenous bulk materials, the neighboring field profiles with opposite phases are identical in amplitude and shape along the boundary, giving the possibility of complete destructive interference with each other in the far-field. However, the only condition preventing them from a complete cancellation is $k_1^x \leq k_2$, which means the mode is above the light line of material 2. Therefore, even for a mode with identical profile of neighboring fields (which is the case of symmetry-protected BICs [13]), the mode has to be below the light line to be bounded, which is clearly contradictory to our current understanding of BIC modes [12, 13].

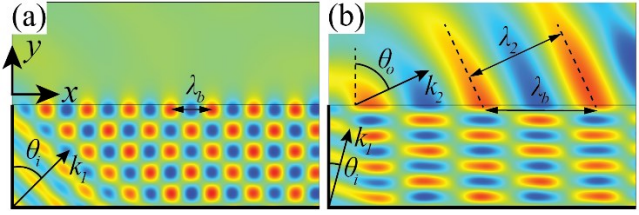

FIG. 1. Light incident from a silicon slab (bottom, material 1) to the vacuum (top, material 2) with an incident angle of (a) $\theta_i = 45°$ and (b) $\theta_i = 15°$.

Here, we discuss the source of the contradictions regarding the concept of light line in photonic band structures, leading to several misunderstandings including BIC modes. To understand the source of this problem, we refer to the original methodology of deriving the photonic band structures [3], in which the authors adopted the standard Bloch theorem from the electronic band structures to the photonic band structures. For an electron in a Bloch state we have

$$\psi = e^{ik.r} u \quad (1)$$

where $\psi$ is the wave function, $u$ is a periodic function with the periodicity of the potential, $r$ is position, and $k$ is the crystal momentum vector. For the state of electrons in crystalline solids, the crystal momentum is restricted within the reciprocal lattice vector $K$, implying that the momentum of an electron in a discrete lattice is given as $k+mK$ for any arbitrary $m$ where $m \in \mathbb{Z}$. This is resulted from the discrete nature of the crystalline solids, which consist of a set of discrete atoms. Electrons oscillate around each of these atoms, forming an electronic wave in the crystal. Therefore, the electronic wave can be considered as a set of discrete oscillating points. In such a system, there are infinite number of discrete wavelengths (momentums) which can fit the phase distribution of those discrete points, governed by the spatial Nyquist frequency which corresponds to the (shortest wavelength) largest momentum supported by the structure, $K$. We can describe this concept in an intuitive way as follows. By increasing the energy of the electrons, the crystal momentum increases until it reaches the value of the reciprocal lattice vector $K$. Beyond $K$, as the energy continues to increase, the crystal momentum decreases until it becomes zero again. Such a cycle repeats as the energy continues to increase. This is the basis of the electronic band structure, in which the energy-momentum relationship can be described completely in the first Brillouin zone by applying the folded bands to each mode. As a result, each crystal momentum corresponds to an infinite number of discrete energies. In this case, the purely-folded nature of the bands is the true representation of the band structure which is determined by the discreteness of the system.

Unlike the discrete electronic periodic structures, photonic structures such as the photonic crystals are continuous, in which the photons can exist at any point of the structure. Therefore, a finite spatial Nyquist frequency doesn't exist. Hence, the wavevector (momentum) of the photons increases continuously as their frequency (energy) increases, resulting in a continuous band structure. In such a continuous band structure, the position of the modes relative to the light line of the surrounding material is essentially different from that of a purely-folded band structure. This flaw of the photonic band structure leads to some misconceptions based on the position of the modes in the band structure, such as the BIC modes determined by the position of high quality-factor modes relative to the light line.

However, a purely-continuous model of band structure may mistakenly imply that a leaky mode lies below the light line on the continuous band, similar to a bound mode. To give a complete picture of the photonic band structure, we propose a continuous-folded model of band structure, in which the continuous and folded bands coexist. In this model, any mode has a component in the continuous band which can be either below or above the light line, as well as a component in the folded band which is above the light line, except near the first folding point. For the frequencies where a mode is bounded, such as a BIC mode, the above-light-line component in the folded band doesn't carry any energy and the whole energy of the mode is in the continuous band below the light line, resulting in a large quality factor. In the case of leaky modes, however, depending on the magnitude of quality factor, part of the mode's energy is in the folded band above the light line with the rest of the energy in the continuous band below the light line. The other case of a leaky mode is that the continuous band itself can be above the light line. As the quality factor of the mode increases, the mode's energy shifts from the folded band to the continuous band below the light line. In contrast, as the quality factor of the mode decreases, the mode's energy shifts from the continuous band below the light line to the folded band.

We show the difference between purely-folded and continuous-folded band structures via a classic design which was used for the experimental demonstration of the trapped light within the radiation continuum [11]. The original band structure in [11] is shown in Fig. 2(b). Here, for simplicity we only plot the band structure from Γ ($k_x=0$, $k_y=0$) to X ($k_x=0.5$, $k_y=0$). The continuous-folded band structure is shown in Fig. 2(a), in which the marker size shows the energy of the mode at each band. The sizes of the continuous and folded bands show the stored and radiated energy after the light propagating for an arbitrary number of wavelengths within the structure (we considered 20,000 wavelengths), respectively, which are calculated based on the quality factor. The high quality-factor TM-like mode near $k_x = 0.27$ is classified as a resonance-trapped BIC in [11]. Based on the conventional purely-folded band structure in Fig. 2(b), this high quality-factor mode is above the light line of background medium (glass), hence should be identified as BIC mode. However, the continuous-folded band structure of the same structure (Fig. 2a) clearly shows that at the high quality-factor point, the above-light-line component vanishes, and the whole energy of the mode is in the below-light-line component. Hence, this mode should be classified as a guided mode.

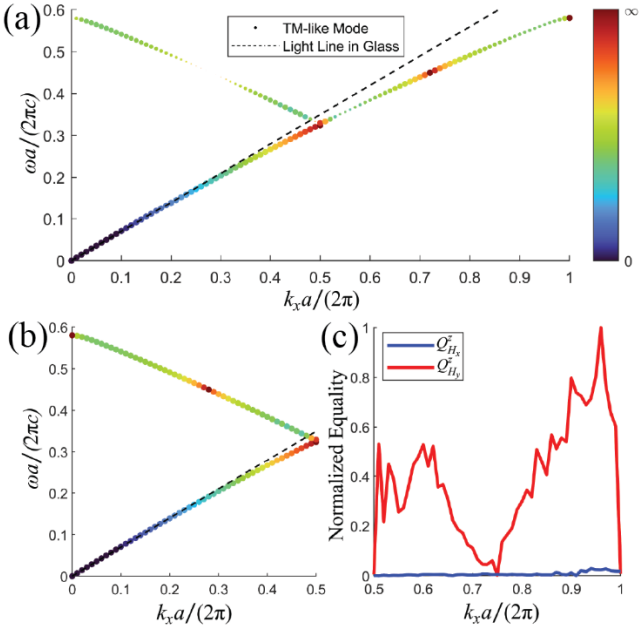

FIG. 2. (a) Continuous-folded and (b) purely-folded band structures as well as (c) the equality of magnetic field components at the boundary of a photonic crystal slab [11]. In (a), the marker size shows the corresponding energy of the bands. In (a) and (b), colormap shows the quality factor of modes.

In order to clarify the underlying physics of boundedness, we introduce two conditions for a bound mode. A mode at a specific frequency needs to satisfy the following two conditions to be bounded in a photonic structure:
1. The mode's continuous band is below the light line.
2. The energy of the mode in the folded band approaches zero.

We expand the second condition to better understand the reason for the existence of folded bands in our continuous-folded band structure model. Over the boundary of a photonic structure, for each field component contributing to the radiation, the interaction of neighboring field profiles with opposite phases may have a residue in the far-field, forming a leaky mode propagating out of the structure. This leaky portion of the mode corresponds to the above-light-line band, and its energy corresponds to the residue of interaction between opposite phases in the far-field. However, when the absolute value of the sum of negative and positive fields is equal to zero, total destructive interference occurs, leading to the almost zero energy in the folded band. In this case, the whole energy is in the below-light-line continuous band.

For a photonic structure radiating in a certain direction, only the field components perpendicular to the radiation direction contribute to the radiation. E.g. to study the radiation along $z$ direction, we only need to consider $E_x, E_y, H_x$ and $H_y$. Furthermore, some of these field components may be absent according to the polarization of light (eg. TE or TM). To study the boundedness of a mode with a radiation channel in $z$ direction, the second condition of boundedness can be presented in the form of the equality of opposite phases of each field corresponding to this radiation channel as

$$Q_F^z = \left| \int_{y_0}^{y_\infty} \int_{x_0}^{x_\infty} F \, dx \, dy \right| \qquad (2)$$

where $F$ is the complex field component, $x_0$ and $x_\infty$ as well as $y_0$ and $y_\infty$ are the lower and upper boundaries of the integration area along $x$ and $y$ axes, respectively. By replacing $F$ with a particular field component such as $H_x$, we can derive the equality of that field component along $z$ direction, such as $Q_{H_x}^z$. Although we only consider $z$ direction here, Eq. (2) can be applied to any direction and its corresponding field components. Because the result of the double integral in Eq. (2) is a complex number, Eq. (2) calculates its absolute value via $|\cdot|$. The value of $Q_F^z$ defines the value of the residual field in the far-field after interference between the opposite phases of the field. For a bound mode, Eq. (2) gives a value approaching zero, implying a total destructive interference in the far-field and the vanishing of the folded band at that frequency.

The boundaries of the integration in Eq. (2) should be defined in a way to cover complete spatial wavelengths of the mode along $x$ and $y$ directions ($\lambda_x = 2\pi/k_x$ and $\lambda_y = 2\pi/k_y$), resulting in $x_\infty - x_0 = \lambda_x$ and $y_\infty - y_0 = \lambda_y$. If the wavevector along any of these directions, such as $x$, is zero (uniform spatial phase along that direction), the integration along that specific direction only needs to cover one period of the photonic crystal along that specific direction such as $x_\infty - x_0 = a$ where $a$ is the periodicity along $x$.

To verify the conditions of boundedness in terms of Eq. (2), we calculate the equalities of the example in Fig. 2. For the TM-like mode, we calculate the equalities of $H_x$ and $H_y$ for radiation in $z$ direction. As shown in Fig. 2(c), $Q_{H_x}^z$ is small for all the values of $k_x$ while $Q_{H_y}^z$ shows valleys at $k_x$=0.5, 0.73, 1, leading to the conclusion that modes at $k_x$=0.5, 0.73, 1 satisfy our second condition of boundedness. Considering that all modes in Fig. 1(a) satisfy the first condition of boundedness, the modes at $k_x$=0.5, 0.73, 1 should be bound modes according to our conditions of boundedness, perfectly matching the computed quality factors as well as the vanishing folded band in Fig. 2(a).

Here, we apply our boundedness conditions to the case of a symmetry-protected mode at $k_x$= 0.5 (Fig. 3). Because the band structure at $k_x$= 0.5 doesn't have a folded band, the continuous band is the only existing band and our second condition of boundedness is satisfied. Because the existence of symmetry-protected BIC doesn't depend on the geometrical parameters of the structure as long as the mirror symmetry of the structure is maintained [13], we are able to manipulate the design parameters without breaking the symmetry condition. In this design example, a 1D periodic structure made of silicon strips on silica substrate, with a periodicity of $a$ =1073 nm and a thickness of $h$=340 nm is considered (Fig. 3a). We consider two designs with different strip widths of $w$ = 142 nm (slim design, Fig. 3e) and $w$ = 222 nm (wide design, Fig. 3f). The structure only has two loss channels via the out-of-plane radiations upward to the air and downward to the substrate. The structure is excited by a TE-polarized light at a wavelength of 1550 nm.

According to the band structure of the slim design, the continuous band is above the light line of glass from $k_x$=0 to $k_x$=0.5 (inset of Fig. 3c), and despite its symmetry-protected mode profile at $k_x$=0.5, it shows a low quality-factor at this point (Fig. 3b). The conventional viewpoint of BICs cannot explain the low quality-factor of a symmetry-protected mode at the highly symmetrical point $k_x$=0.5 in the slim design. However, according to our conditions of boundedness, it is clear that the quality factor of the slim design shown in Fig. 3(b) stays low because the continuous band of the mode is above the light line (breaking our first condition of boundedness).

For the wide design, the mode is above the light line similar to the slim design, however it goes below the light line near $k_x$= 0.5 (inset of Fig. 3d), which causes the quality factor to be increased sharply by a factor of $10^7$ (Fig. 3b). According to our first condition of boundedness, the bound mode at $k_x$= 0.5 is only caused by the fact that continuous band goes below the light line, rather than the change in the symmetry condition of the mode. This example verifies our first condition of boundedness and demonstrates that our proposed band structure gives the correct interpretation of the boundedness of modes.

This work reveals a fundamental flaw of photonic band structures. This flaw originates from the adoption of purely-folded band structure from electronic crystals to photonic crystals without considering an essential difference between electronic and photonic crystals — electronic crystal is discrete while its photonic counterpart is continuous. We proposed that continuous systems such as photonic crystals should be characterized by the continuous-folded band structures. We showed that in the continuous-folded band structures, all the bound modes are below the light lines of the surrounding medium. We proposed two conditions for determining the boundedness of photonic modes. Our conditions successfully confirmed the existence of a resonance-trapped bound mode, as well as the quality factor of a mode at a highly symmetrical point by determining the position of the continuous band relative to the light line. The quality factor of this mode, however, cannot be correctly predicted by the principle of conventional symmetry-protected BIC.

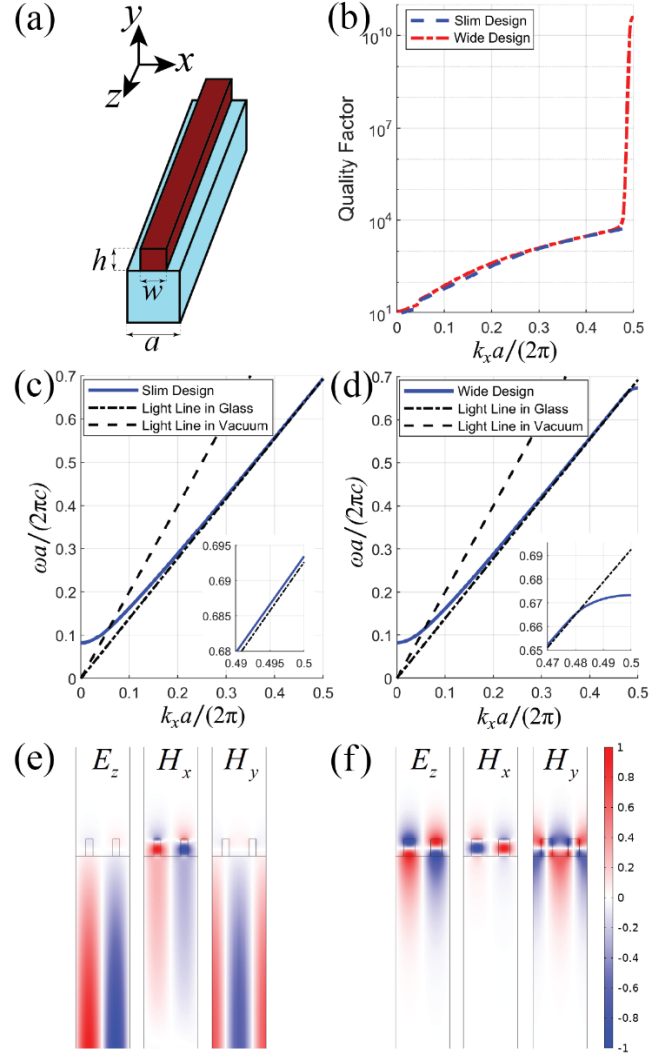

FIG. 3. Structure, quality factors, band structures and field profiles of a 1D photonic crystal. (a) Schematic of the photonic crystal. (b) The quality factors of the slim and wide designs. The band structures of the (c) slim and (d) wide designs (the insets show the magnified region near $k_x$ approaching 0.5). The field profiles of (e) slim and (f) wide designs at $k_x$=0.5.

The author would like to thank Yang Li and Yuanmu Yang for the discussions as well as helpful comments and suggestions. We acknowledge support from the National Natural Science Foundation of China (62075114) and Beijing Natural Science Foundation (4212050). This work is supported by the Center of High-Performance Computing, Tsinghua University.